\documentclass[reqno]{amsart}
 \usepackage{amsbsy,amssymb,amscd,amsfonts,latexsym,amstext,delarray,
 amsmath,graphicx, color}
 \usepackage[dvips]{epsfig}
\input xypic
\usepackage{hyperref}

\definecolor{Green}{rgb}{0.0,0.40,0.0}

\newtheorem{thm}{Theorem}[section]

\def\qqq{\,,\quad~\forall}

\def\Aut{{\rm Aut}}

\def\Diff{{\rm Diff}}

\def\SU{{\rm SU}}
\def\Spec{{\rm Spec}}
\def\Sp{{\rm Spec}}

\def\Tr{{\rm Tr}}

\def\N{{\mathbb N}}

\def\R{{\mathbb R}}
\def\Z{{\mathbb Z}}

\def\Tr{{\rm Tr}}

\def\cA{{\mathcal A}}
\def\cB{{\mathcal B}}

\def\cG{{\mathcal G}}
\def\cH{{\mathcal H}}

\def\cL{{\mathcal L}}

\newcommand{\ie}{{\it i.e.\/}\ }
\newcommand{\eg}{{\it e.g.\/}\ }
\newcommand{\cf}{{\it cf.\/}\ }

\def\Int{{\mbox{Int}}}

\newcommand{\nil}[1]{}

\title
{Space-Time from the spectral point of view}

\date{}
\begin{document}
\maketitle

\centerline{\large\bf Ali H.
Chamseddine$^{1,3}$\ , \ Alain Connes$^{2,3,4}$\ \ } \vspace{.5truecm}
\emph{\centerline{$^{1}$Physics Department, American University of Beirut, Lebanon}}
\emph{\centerline{$^{2}$College de France, 3 rue Ulm, F75005, Paris, France}}
\emph{\centerline{$^{3}$I.H.E.S. F-91440 Bures-sur-Yvette, France}}
\emph{\centerline{$^{4}$Department of Mathematics, Vanderbilt University, Nashville, TN 37240 USA}}

\vspace{2cm}

\begin{abstract} We develop the spectral point of view on geometry based on the formalism of quantum physics. We start from the simple physical question of specifying our position in space and explain how the spectral geometric point of view provides a  new paradigm to model space-time whose fine structure can be encoded by a finite geometry. The classification of the irreducible finite geometries of $KO$-dimension $6$ singles out a  ``symplectic--unitary" candidate $F$, which when used as the fine texture of space-time delivers from pure gravity on $M\times F$ the Standard Model coupled to gravity and, once extrapolated to unification scale, gives testable predictions.
\end{abstract}

\tableofcontents

\section{Introduction} Let us start with the following simple question:

\centerline{``How do we tell where we are in space"}

\begin{figure}
\begin{center}
\includegraphics[scale=1]{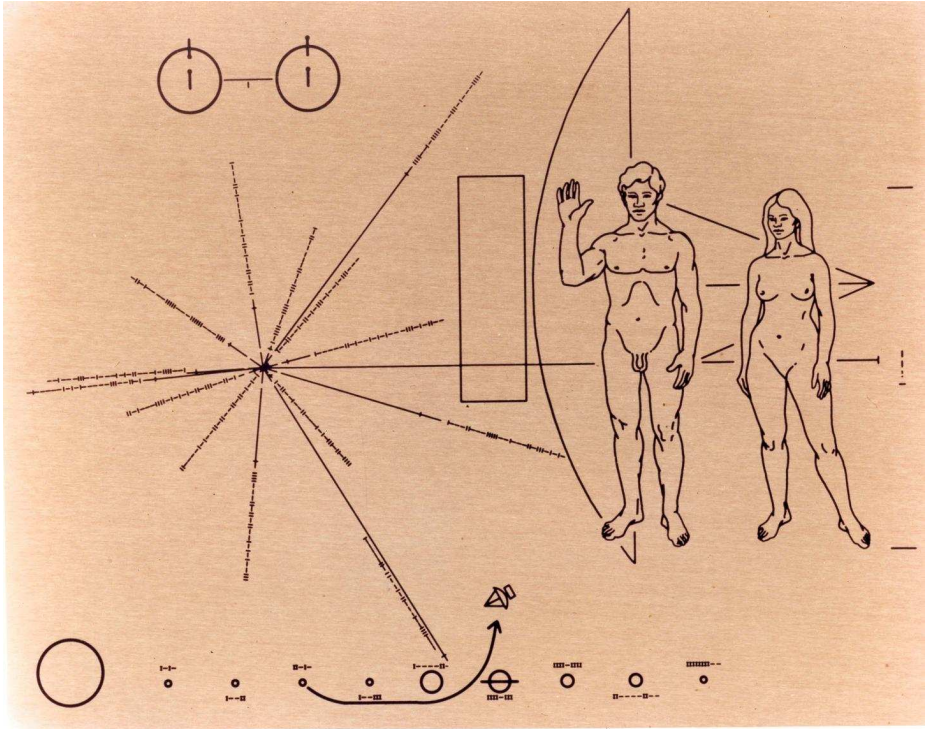}
\end{center}
\caption{The pioneer plaque, besides the binary relative distances
of the planets of the solar system it gives the position of the sun relative to
$14$ pulsars and the center of the galaxy.\label{probe} }
\end{figure}

An example of a possible answer is displayed in Figure \ref{probe}, which represents the pioneer plaque \ie the message of the pioneer probe. Besides a picture of human beings and of the outer appearance of the solar system it contains the position of the sun relative to
$14$ pulsars and the center of the galaxy. It also gives in the upper left corner the hyperfine transition of neutral hydrogen.

From a geometric stand point it is natural to specify our position $x$ in space by giving curvature invariants at $x$, under the standard hypothesis that space is well modeled as a Riemannian space $X$ of dimension three. Of course there is no way to distinguish between points which are obtained from each other by an isometry of $X$. A very related problem is the problem of giving observable quantities in the theory of gravity. An observable should be an invariant of the geometry.

Besides the above geometric point of view there is a ``dual" one which is based on spectral invariants and whose relation to the geometric one is through the heat kernel expansion of the trace of operators in Hilbert space. Our thesis is that, since much of the information we have about the nature of space-time is of spectral nature, one needs to understand carefully the process which transforms this spectral information into a geometric one. At the mathematical level one knows that the spectrum of the Dirac operator $D$ of a compact Riemannian space gives a sequence of invariants of the geometry: the list of the eigenvalues. It is also known from Milnor's one page paper \cite{Milnor}, that this invariant is not complete. The missing information is given by the relative position in Hilbert space of two commutative algebras $A$ and $B$. The first is the algebra of measurable bounded functions on $X$ acting by multiplication in the Hilbert space $\cH$ of $L^2$-spinors. It is an old result of von Neumann that, once the dimension of $X$ is fixed, the pair $(A,\cH)$ does not depend upon the manifold $X$. Said in simple terms what it means is that, from the point of view of measure theory, the ``bunches of points" coming from different manifolds of the same dimension can be identified and say nothing about the geometry. Similarly the spectral theory of operators tells us that the spectrum of the Dirac operator $D$ \ie its list of eigenvalues as a subset with multiplicity inside $\R$ gives us the full information about the pair $(\cH,D)$ of the Hilbert space of $L^2$-spinors and the Dirac operator acting in $\cH$. The missing invariant (\cf \cite{CKM}) that one needs in order to assemble together the pieces $(A,\cH)$ and $(\cH,D)$ and obtain the spectral triple $(A,\cH,D)$ is given by the relative position of the algebra $A$ with the algebra $B$ of functions $f(D)$ of the Dirac operator. The  invariant defined in \cite{CKM} is an infinite dimensional analogue of an invariant which is familiar to physicists and which measures the relative position of the mass eigenstates for the upper  quarks with respect to the mass eigenstates for the lower ones lifted up using the action of the weak isospin group. This is the CKM matrix which measures the generalized ``angle" of two basis (whose elements are given up to phase). Once the spectral triple $(A,\cH,D)$ is assembled from its two pieces one recovers the points of the space and the full geometric information. It is interesting that the invariant manner of encoding a point $x$ in this process can be understood by specifying an infinite hermitian matrix $H_{\lambda\mu}$ with complex entries. The labels $\lambda,\mu$ are the eigenvalues of the Dirac operator $D$. The matrix elements $H_{\lambda\mu}$ are the inner products
$$
H_{\lambda\mu}=\langle \psi_\lambda(x),\psi_\mu(x)\rangle
$$
of the eigenspinors $\psi_\lambda$ evaluated at the point $x$. While these eigenspinors are globally
orthogonal, they are not so when evaluated at a point $x$ since the spinor space $S_x$ at $x$ is finite dimensional. Thus the matrix $H_{\lambda\mu}$  expresses the correlations between different frequencies at the point $x$. Modulo the obvious gauge ambiguity this matrix characterizes the point $x$ (\cf \cite{bbg}, \cite{CKM}) in a way which is dual to the local expansion of curvature invariants of the metric. This suggests that in order to specify ``where we are" we should first give the spectrum of the Dirac operator and then the matrix $H_{\lambda\mu}$ which, in essence, gives the correlations between the various frequencies. It is not quite what happens concretely in physics since most of the observations which are done involve light (bosons), rather than fermions such as neutrinos, but we are nevertheless quite used to the need for labeling the information in a directional manner (as what would happen using spinor space at $x$) and for establishing correlations between observations at different frequencies  such as infrared and ultraviolet ones.

In this short survey paper we shall explain how the operator formalism of quantum mechanics provides all the needed tools to reconstruct the geometry from the spectral data and at the same time suggests that the algebra of coordinates on space-time possesses a small amount of non-commutativity which is responsible for the three other forces besides gravity.

\section{The quantum variability}

The two notions of ``variable" and of ``infinitesimal" played a central role at the very beginning of the calculus.
According to Newton:
{\em ``In a certain problem, a variable is the quantity that takes an infinite number of values which are quite determined by this problem and are arranged in a definite order"}

Moreover he explicitly defined and considered  infinitesimal variables:
{\em ``A variable is called infinitesimal if among its particular values one can be found such that this value itself and all following it are smaller in absolute value than an arbitrary given number"}

In our modern language we are used to think of  a real variable as a map
$$
f:X\to \R
$$
from a set $X$ to the real line $\R$. Let us start from the remark that discrete and continuous variables cannot coexist in this  formalism. The simple point is that if a variable is continuous the set $X$ is necessarily at least of the cardinality of the continuum and this precludes the existence of a variable with countable range such that each value is reached only a finite number of times. This problem of treating continuous and discrete variables on the same footing is  solved using the formalism of quantum mechanics. In this formalism a ``real variable" is not given in the above classical manner but is a self-adjoint operator in Hilbert space. As such it has a ``spectrum" which is its set of values, each being reached with some (spectral) multiplicity. A continuous variable is an operator with continuous spectrum and a discrete variable an operator with discrete spectrum (and of course the mixed case occurs).   The uniqueness of the separable infinite dimensional Hilbert space  shows that the Hilbert space  $L^2[0,1]$ of square integrable functions on the unit interval, is the same as the Hilbert space of square integrable sequences $\ell^2(\N)$.  This shows that variables with continuous range, such as $T$ where $(T\xi)(x)=x\xi(x)$ for $\xi\in L^2[0,1]$,  coexist  with variables with countable range such as $S$, $(Sa)_n=\frac 1na_n$ for $(a_n)\in \ell^2(\N)$. The only new fact is that they just cannot commute. If they would it would have been possible to model them in a classical manner but this is not the case.
In classical physics the basic variability is due to the passing of time so that ``t" is the paradigm of the ``variable". But in quantum physics there is a more profound  inherent {\em variability} which is  a basic experimental physics fact. It prevents one from reproducing experimental results of quantum physics which display the choice of an eigenvalue by reduction of the wave packet, as in the diffraction of an electron by a narrow slit. This spontaneous variability of the quantum world far surpasses in originality the simple time variation of the classical world.

In the formalism of quantum mechanics there is a reserved place for the ``infinitesimal variables" which correspond exactly to Newton's definition, they are such that for any $\epsilon>0$ one can find an eigenvalue of the absolute value such that this value itself and all following it are smaller
than $\epsilon$. These operators are called ``compact" and they satisfy all the algebraic relations that one would naively expect from infinitesimals (in particular they form a two-sided ideal). We shall not recall here the dictionary that translates from classical concepts to the quantum ones, the most original one being the way one integrates infinitesimals of order one, by picking the coefficient of the logarithm $\log(n)$ in the sum of the first $n$-eigenvalues of the operator. We refer to \cite{Co-book} Chapter IV for the full treatment.

\section{From the Riemannian paradigm to the spectral one}

The Riemannian paradigm is based on the Taylor expansion in local coordinates
of the square of the line element and in order to measure the distance between
two points one minimizes the length of a path joining the two points
$$
d(a,b)=\,{\rm Inf}\,\int_\gamma\,\sqrt{g_{\mu\,\nu}\,dx^\mu\,dx^\nu}
$$
Thus in this paradigm of geometry only the square of the line element makes sense, and the formula for the geodesic distance involves the extraction of a square root. This extraction of a square root is in fact hiding a deeper understanding of the line element and the choice of a square root is associated to a global structure which is that of a spin structure. The Dirac operator $D$ is canonically associated to a Riemannian metric and a spin structure by a formula which can be traced back to Hamilton (who wrote down the operator $i\partial_x+ j\partial_y+k\partial_z$ in terms of his generators $i,j,k$ for quaternions).
Paul Dirac showed, in the flat case, how to extract the square root of the Laplacian in order to obtain a first order version of the Maxwell equation and Atiyah and Singer gave the general canonical definition of the Dirac operator on a Riemannian manifold endowed with a spin structure. This provides a direct connection with the quantum formalism: the line element is now upgraded to this formalism as the propagator
$$
ds=D^{-1}
$$
and the same geodesic distance $d(a,b)$   can be computed in a dual manner as
$$
d(a,b)=\,{\rm Sup}\, \{ \vert f(a) - f(b) \vert \,;\, f \in {
A}\, , \ \Vert [D,f] \Vert \leq 1\,\}
$$
where, as above, $A$ is the algebra of measurable functions acting by multiplication in the Hilbert space $\cH$ of $L^2$-spinors. In other words one measures distances not by taking the shortest continuous path between the two points $a,b$, but by sending a wave $f(x)$ whose frequency is limited from above, and measuring its maximal variation $\vert f(a) - f(b) \vert $ from $a$ to $b$. The operator norm $\Vert [D,f] \Vert$ controls from above the frequency of $f$ since it is given by the supremum of the gradient of $f$ measured using the Riemannian metric. It is quite important to understand at this point how one reconstructs the underlying space $X$. A point $a$ of $X$ is a character  $f\mapsto f(a)$ of the subalgebra of $A$ given by the condition $f \in {
A}\, , \ \Vert [D,f] \Vert \leq \infty$. This condition involves the choice of $D$ and determines the Lipschitz functions inside the algebra of measurable functions. Thus in particular the operator $D$ determines what it means to be continuous, and hence the topology of $X$. In fact it also determines smoothness and the latter comes from the one parameter group $e^{it |D|}$ which plays the role of the geodesic flow in the operator theoretic framework. The algebra $A$ is fixed once and for all thanks to the following Theorem of von Neumann
(\cite{vNeumann}, \cite{vNeumann1})  which shows that
there is a unique way to represent the continuum with constant multiplicity $m$.

\begin{thm}\label{vNthm}
Let $\cH$ be an infinite dimensional Hilbert space  with countable
orthonormal basis  and $m $ an integer. There exists up to
unitary equivalence only one commutative von Neumann subalgebra $A\subset
\cL(\cH)$ such that,
\begin{enumerate}
  \item $A$ contains no minimal projection,
  \item The commutant of $A$ is isomorphic to $M_m(A)$.
\end{enumerate}
\end{thm}

This Theorem should be thought of as the operator theoretic version of the uniqueness of the continuum as a {\em set}.

 It  shows that if we are interested in sifting through all Riemannian geometries of a given dimension, we can fix the algebra $A$ and the way it is represented as operators in the Hilbert space $\cH$ (which is unique itself also).

  Thus the only remaining ``variable" is the operator $D$. This operator has two invariants which are also invariants of the geometry:
\begin{itemize}
  \item Its spectrum $\Spec\,D$.
  \item The relative spectrum $\Sp_N(A)$ where $N=\{f(D)\}$ is the algebra of functions of $D$.
\end{itemize}
We refer to \cite{CKM} for the definition of the relative spectrum, which as explained in the introduction measures the relative position of the two algebras $A$ and $N$.

\section{Why go noncommutative}
In the above discussion of the spectral point of view on ordinary geometry the noncommutativity of the infinitesimal line element $ds$ with the coordinates played a basic role but the algebra of coordinates was still commutative. We shall now explain why it is important that the spectral point of view is tailor-made to treat noncommutative algebras of coordinates on the same footing as the commutative ones.

The geometric understanding of gravitation provided by the general theory of relativity is sufficiently compelling that it is very desirable to try and incorporate the other three forces in the same scheme. This has been tried by increasing the dimension of the space-time manifold since the work of Kaluza and Klein. The virtue of the noncommutative approach is that it resolves an issue that immediately arises in the Kaluza-Klein models.
The point is that the group $\cG$ of symmetries of the Lagrangian of gravity coupled with matter is handed to us by physics. It is the semi-direct product of the group ${\rm Map}(M,G)$  of gauge transformations of second kind (\ie maps from the manifold $M$ to the small gauge group $G$) by the symmetry group of gravity, namely the group $\Diff(M)$ of diffeomorphisms of ordinary space-time $M$:
$$
\cG={\rm Map}(M,G)\ltimes
\Diff(M)
$$
This decomposition   is similar to the decomposition of the Poincar\'e group as a semi-direct product of the subgroup of translations by the Lorentz group.
Now if gravity coupled with matter is going to be pure gravity on a new space $N$, and unless we want to argue that a different symmetry group is hidden behind physics to be discovered which is wishful thinking,  the most obvious requirement is to find the manifold $N$ in such a way that
\begin{equation}\label{solve}
\Diff(N)=\cG
\end{equation}
So one can browse through books computing diffeomorphism groups of higher dimensional manifolds $N$ and hope for the best. The trouble is that there is no solution. This comes from a general mathematical result which asserts that the connected component of identity in $\Diff(N)$ is a {\em simple} group for any manifold $N$. Thus, since  $\cG$ has the non-trivial normal subgroup ${\rm Map}(M,G)$ there is no way one can solve the above equation \eqref{solve} using ordinary manifolds $N$.
Let us now show that \eqref{solve} admits a solution, \ie that the group $\cG$ is indeed the group of diffeomorphisms of a new space $N$, provided  one searches for noncommutative solutions. The small gauge group will be $G=\SU(n)$ divided by its finite center. In fact, to obtain the solution,  it will be enough to perform on the original space $M$ a very simple operation which consists in replacing its algebra of coordinates by the algebra of matrices over it.
If we start with an algebra $\cA$, we get for each integer $n$ a new algebra $M_n(\cA)$ by considering matrices with entries in $\cA$ which we add and multiply as ordinary matrices, thus for $n=2$ we get
$$
\left(\begin{array}{cc}
 a_{11} & a_{12} \\
 a_{21} & a_{22} \\
  \end{array}
   \right)\,
   \left(\begin{array}{cc}
 b_{11} & b_{12} \\
 b_{21} & b_{22} \\
  \end{array}
   \right)\,=
   \left(\begin{array}{cc}
 a_{11} b_{11}+a_{12}b_{21}& a_{11} b_{12}+a_{12}b_{22} \\
 a_{21} b_{11}+a_{22}b_{21} & a_{21}b_{12} +a_{22} b_{22}\\
  \end{array}
   \right)\,
$$
for the product of two elements and this only uses the rules of addition and multiplication in $\cA$. It is obvious that even if $\cA$ was commutative to start with, the new algebra $M_n(\cA)$ is no longer commutative as soon as $n>1$. Let $\cA=C^\infty(M)$ be the   algebra
of smooth  functions on a manifold $M$. We take complex valued functions and remember the real valued ones by the natural involution on $\cA$ given by $f^*(x)=\overline{f(x)}$ for all $x\in M$.

It is then a simple fact that the group of diffeomorphisms of $M$ can be expressed algebraically as the group of automorphisms of $\cA$,
 $$
 \Diff(M)=\Aut(\cA).
 $$
 To a diffeomorphism $\varphi$ corresponds the automorphism $\theta$ given by the formula
 $$
 \theta(f)(x)=f(\varphi^{-1}(x))\qqq x\in M
 $$
 Let us now consider the new algebra $\cB=M_n(\cA)$ and compute $\Aut(\cB)$ which plays the role of  $\Diff(N)$ for the noncommutative space whose algebra of coordinates is $\cB$.
 Since $\cB$  is noncommutative, unitary elements $u\in \cB$, $uu^*=u^*u=1$ define  automorphisms of $\cB$ by the equality
$$
{\rm Ad}(u)(x)=uxu^*\qqq x\in \cB
$$
These automorphisms are non trivial unless $u$ is in the center of $\cB$ and they form a normal subgroup $\Int(\cB)$ of $\Aut(\cB)$. This shows very generally that the group of automorphisms of a noncommutative algebra $\cB$ admits a normal subgroup, the group $\Int(\cB)$ of inner or ``internal" automorphisms. In the specific case considered above, \ie $\cB=M_n(C^\infty(M))$, one gets
\begin{equation}\label{semi}
\Aut(\cB)={\rm Map}(M,G)\ltimes \Diff(M)
\end{equation}
where the small gauge group $G$ is the group $\SU(n)$ divided by its finite center. Thus this gives a solution to equation \eqref{solve}.

 We have shown in  \cite{cc2} that the study of pure gravity for spectral geometries involving the algebra   $\cB=M_n(C^\infty(M))$
 instead of the usual commutative algebra $C^\infty(M)$ of smooth functions,  yields
Einstein gravity on $M$ minimally coupled with Yang-Mills theory for the gauge
group $\SU(n)$. The Yang-Mills gauge potential appears as the inner part of the
metric, in the same way as the group of gauge transformations (for the gauge
group $\SU(n)$) appears as the group of inner diffeomorphisms. This simple example shows that the noncommutative world incorporates the internal symmetries in a natural manner as a slight refinement of the algebraic rules on coordinates. There is a certain similarity between this refinement of the algebraic rules and what happens when one considers super-space in supersymmetry, but unlike in the latter case the algebraic rules are semi-simple rather than nilpotent. The effect is also somewhat similar to what happens in the Kaluza-Klein scenario since it is pure gravity on the new geometry that produces the mixture of gravity and gauge theory. But there is a fundamental difference since the construction does not alter the metric dimension and thus does not introduce the infinite number of new modes which automatically come up in the Kaluza-Klein  model. In this manner one stays much closer to the original input from physics and does not have to argue that the new modes are made invisible because they are very massive.

\section{How to go noncommutative}

The spectral point of view which is based on operators acting in Hilbert space and on the formalism of quantum mechanics, is ready made for noncommutative algebras of coordinates.

\bigskip

\begin{center}
\begin{tabular}{|c|c|}
\hline &  \\
Space $X$ & Algebra \ ${\mathcal A}$
\\
&\\
\hline &  \\
 Real variable & Self-adjoint  \\
$x^\mu$ & operator  $H$\\
&\\
\hline &  \\
 Infinitesimal &  Compact  \\
 $dx$ & operator $\epsilon$ \\
&\\ \hline &  \\
 Integral of  & $\int \epsilon =$ coefficient of \\
 infinitesimal & $\log(\Lambda)$ in $\Tr_\Lambda$($\epsilon$)\\
&\\ \hline & \\
 Line element  & $D^{-1}=$ Fermion \\
$\sqrt{g_{\mu\nu}\,dx^\mu dx^\nu}$& propagator \\
  &\\ \hline
\end{tabular}

\end{center}

\bigskip

Thus, the basic data is that of a spectral triple $(\cA,\cH,D)$ which gives a representation in Hilbert space $\cH$ of both the algebra $\cA$ of coordinates and of the inverse line element $D$.
In order to cope with noncommutativity of $\cA$, one needs to understand a very fundamental result which is due to the Japanese operator algebraist M. Tomita \cite{TT}. What it says is that under fairly general circumstances one can, given a von Neumann algebra $A$ of operators in Hilbert space $\cH$, find an antiunitary  isometry $J$ such that the following commutators vanish:
$$
[x,Jy^*J^{-1}]=0\qqq x,y\in A.
$$
In other words, even though $A$  is noncommutative there is still commutativity around, namely
$$
xy^{0}=y^{0}x\qqq x,y\in A,\ \ y^{0}=Jy^*J^{-1}.
$$
It is at this point that one witnesses an amazing confluence between the theory of operator algebras and the formalism of $K$-theory \cite{Atiyah}. Indeed while at first sight one might attribute the non-trivial global properties of manifolds to ordinary Poincar\'e duality, the work of geometers in the seventies has shown that more refined invariants such as the Pontrjagin classes witness Poincar\'e duality at a deeper level, that of $KO$-theory. Moreover, as a byproduct of the index theorem of Atiyah-Singer a purely operator theoretic formulation of the  $KO$-homology cycles was obtained (\cf \cite{Atiyah1}, \cite{Kas}). The confluence which is a kind of birth certificate for noncommutative manifolds is that, besides the elements $(\cA,\cH,D)$ of the spectral triple, the same antiunitary operator $J$ plays a key role in the formulation of $KO$-homology cycles. The basic rules are
$$
[a,b^{0}]=0\,, \ [[D,a],b^{0}]=0\,, \ b^{0}=Jb^*J^{-1}
$$
$$
J^2=\varepsilon\,, \ DJ=\varepsilon^{\prime}JD,\quad
J\,\gamma=\varepsilon^{\prime\prime }\gamma J,\quad D\gamma=-\gamma D
$$
where $\gamma$ is the $\Z/2$ grading operator which only exist in the even dimensional case
and anticommutes with the operator $D$. The $KO$-theory comes in $8$ different versions which just depend upon the dimension of the geometry modulo $8$. They are distinguished by the three possible signs $\epsilon\in \pm 1$ which govern the above algebraic rules and whose values according to the dimension modulo $8$ are:

\bigskip

\begin{center}
 \begin{tabular}
[c]{|c|rrrrrrrr|}\hline \textbf{n } & 0 & 1 & 2 & 3 & 4 & 5 & 6 &
7\\\hline\hline
$\varepsilon$ & 1 & 1 & -1 & -1 & -1 & -1 & 1 & 1\\
$\varepsilon^{\prime}$ & 1 & -1 & 1 & 1 & 1 & -1 & 1 & 1\\
$\varepsilon^{\prime\prime}$ & 1 &  & -1 &  & 1 &  & -1 & \\\hline
\end{tabular}
\end{center}

\bigskip

Thus a spectral manifold is given by a spectral triple $(\cA,\cH,D)$
with the further structure provided by the unitary involution $J$ (and in the even case the
$\Z/2$ grading $\gamma$). In physics terms these data have the following names and meaning:
\begin{itemize}
  \item $\cH$: one particle Euclidean Fermions
  \item $D$: inverse propagator
  \item $J$: charge conjugation
  \item $\gamma$: chirality
\end{itemize}
and thus the new formalism for geometry keeps a very close contact with physics. Exactly
as the inner automorphisms form an ``internal" part of the group of geometric symmetries, the
metric admits ``inner fluctuations" and we refer \eg to \cite{CMbook} for a detailed treatment of the latter.

\section{A dress for the beggar}

In our first approach \cite{cc1} to the understanding of the Lagrangian of
the Standard Model coupled to gravity, we used the  above new paradigm of
spectral geometry to model space-time as a  product of an ordinary
$4$-manifold (we work after Wick rotation in the Euclidean signature)  by a
finite geometry $F$. This finite geometry was taken from the phenomenology
i.e. put by hand to obtain the Standard Model Lagrangian using the spectral
action. The algebra $\mathcal{A}_{F}$, the Hilbert space $\mathcal{H}_{F}$ and
the operator $D_{F}$ for the finite geometry $F$ were all taken from the
experimental data. The algebra comes from the gauge group, the Hilbert space
has as a basis the list of elementary fermions and the operator is the Yukawa
coupling matrix. This worked fine for the minimal Standard Model, but there
was a problem \cite{lizzi} of doubling the number of Fermions, and also the Kamiokande
experiments on solar neutrinos showed around $1998$ that, because of neutrino
oscillations, one needed a modification of the Standard Model incorporating in
the leptonic sector of the model the same type of mixing matrix already
present in the quark sector. One further needed to incorporate a subtle
mechanism, called the see-saw mechanism, that could explain why the observed
masses of the neutrinos would be so small. At first our reaction to this
modification of the Standard Model was that it would certainly not fit with
the noncommutative geometry framework and hence that the previous agreement
with noncommutative geometry was a mere coincidence. After about $8$ years it
was shown in  \cite{CoSMneu}  and \cite{mc2} that the only needed change
(besides incorporating a right handed neutrino per generation) was to make a
very simple change of sign in the grading for the anti-particle sector of the
model (this was also done independently in \cite{Barrett}). This not only
delivered naturally the neutrino mixing, but also gave the see-saw mechanism
and settled the above Fermion doubling problem. The main new feature that
emerges is that when looking at the above table of signs giving the
$KO$-dimension, one finds that the finite noncommutative geometry $F$ is now
of dimension $6$ modulo $8$. Of course the space $F$ being finite, its metric
dimension is $0$ and its inverse line-element is bounded. In fact this is not
the first time that spaces of this nature--- i.e. whose metric dimension is
not the same as the $KO$-dimension--- appear in noncommutative geometry and
this phenomenon had already appeared for quantum groups and related homogeneous spaces \cite{dab}.

\smallskip

Besides yielding the Standard Model with neutrino mixing and making testable
predictions (as we shall see in \S \ref{predsect}), this allowed one to hope
that, instead of taking the finite geometry $F$ from experiment, one should in
fact be able to derive it from first principles. The main intrinsic reason for
crossing by a finite geometry $F$ has to do with the value of the dimension of
space-time modulo $8$. We would like this $KO$-dimension to be $2$ modulo $8$
(or equivalently $10$) to define the Fermionic action, since this
eliminates\footnote{because this allows one to use the antisymmetric bilinear
form $\langle J\xi,D\eta\rangle$ (for $\xi,\eta\in\mathcal{A}_{F},\gamma
\xi=\xi,\gamma\eta=\eta$). The appearance of dimension $10$ is for the same
reason as in supersymmetric theories where both Majorana and Weyl conditions
are imposed simultaneously. } the doubling of fermions in the Euclidean
framework. In other words the need for crossing by $F$ is to shift the
$KO$-dimension from $4$ to $2$ (modulo $8$).

\smallskip This suggested to us to classify the simplest possibilities for the
finite geometry $F$ of $KO$-dimension $6$ (modulo $8$) with the hope that the
finite geometry $F$ corresponding to the Standard Model would be one of the
simplest and most natural ones. This was finally done recently (\cite{cc5},
\cite{cc6}).

\smallskip

From the mathematical standpoint our road to $F$ is through the following steps

\begin{enumerate}
\item We classify the irreducible triplets $(\mathcal{A},\mathcal{H},J)$.

\item We study the $\mathbb{Z}/2$-gradings $\gamma$ on $\mathcal{H}$.

\item We classify the subalgebras $\mathcal{A}_{F} \subset\mathcal{A}$ which
allow for an  operator $D$ that does not commute with the center of
$\mathcal{A}$ but fulfills the  ``order one" condition:
\[
[[D,a],b^{0}] = 0 \qquad\forall\, a,b \in\mathcal{A}_{F} \,.
\]

\end{enumerate}

The classification in the first step shows that the solutions fall in two
classes, in the first the dimension $n$ of the Hilbert space $\mathcal{H}$ is
a square: $n=k^{2}$, in the second case it is of the form $n=2k^{2}$. In the
first case the solution is given by a real form of the algebra $M_{k}%
(\mathbb{C})$ of $k \times k$ complex matrices. The representation is given by
the action by left multiplication on $\mathcal{H}=M_{k}(\mathbb{C})$, and the
isometry $J$ is given by $x\in M_{k}(\mathbb{C})\mapsto J(x)= x^{*}$. In the
second case the algebra is a real form of the sum $M_{k}(\mathbb{C})\oplus
M_{k}(\mathbb{C})$ of two copies of $M_{k}(\mathbb{C})$ and while the action
is still given by left multiplication on $\mathcal{H}=M_{k}(\mathbb{C})\oplus
M_{k}(\mathbb{C})$, the operator $J$ is given by $J(x,y)=(y^{*},x^{*})$.

The study (2) of the $\mathbb{Z}/2$-gradings shows that the commutation
relation $J\gamma=-\gamma J$ excludes the first class. Thus, since we want  the
finite geometry $F$ to be of $KO$-dimension $6$, we are left only
with the second case and we obtain among the very few choices of lowest
dimension the case $\mathcal{A}=M_{2}(\mathbb{H})\oplus M_{4}(\mathbb{C})$
where $\mathbb{H}$ is the skew field of quaternions. At a more invariant level
the Hilbert space is then of the form $\mathcal{H}=\mathrm{Hom}_{\mathbb{C}%
}\left(  V,W\right)  \oplus\mathrm{Hom}_{\mathbb{C}}\left(  W,V\right)  $
where $V$ is a $4$-dimensional complex vector space, and $W$ a two dimensional
graded right vector space over $\mathbb{H}$. The left action of $\mathcal{A}%
=\mathrm{End}_{\mathbb{H}}\left(  W\right)  \oplus\mathrm{End}_{\mathbb{C}%
}\left(  V\right)  $ is then clear and its grading as well as the grading of
$\mathcal{H}$ come from the grading of $W$. Note that this determines the
number of fermions to be $4^{2}=16.$

Our main result then is that there exists up to isomorphism a unique
involutive subalgebra of maximal dimension $\mathcal{A}_{F}$ of $\mathcal{A}%
^{\mathrm{ev}}$, the even part\footnote{One restricts to the even part to
obtain an ungraded algebra.} of the algebra $\mathcal{A}$, which solves (3).
This involutive algebra $\mathcal{A}_{F}$ is isomorphic to $\mathbb{C}%
\oplus\mathbb{H}\oplus M_{3}(\mathbb{C})$ and together with its representation
in $(\mathcal{H},J,\gamma)$ gives the noncommutative geometry $F$ which we
used in \cite{mc2} to recover the Standard Model coupled to gravity using the
spectral action which we  describe below in \S \ref{spectralaction}.

This result is remarkable because the input that was used is minimal and the first
possibility obtained consistent with the axioms of noncommutative geometry,
after imposing the symplectic-unitary symmetry condition on the algebra, is
the algebra of the standard model with the fermions in the correct
representation. All the arbitrariness that is usually encountered in the
construction of the standard model whether in the choice of the $SU(3)\times
SU(2)\times U(1)$ gauge group, the fermionic representations, or the Higgs
structure and the electroweak spontaneous breaking mechanism disappear. The
standard model becomes completely determined. In this respect we see that
there is a geometrical structure responsible for all the details of the
standard model. The beggar (SM) is now beautifully  dressed and noncommutative
geometry has revealed the inner beauty of the construction. Geometrically we
see that the underlying algebra is a direct sum of two algebras. The first
algebra is quaternionic $M_{2}\left(  \mathbb{H}\right)  $, broken to $\left(
\mathbb{C}\oplus\mathbb{C}\right)  _{R}\oplus\mathbb{H}_{L},$ decomposes by
the chirality operator into a left-handed and right-handed sectors. The second
algebra $M_{4}\left(  \mathbb{C}\right)  $ is broken into $\mathbb{C}\oplus
M_{3}\left(  \mathbb{C}\right)  $ and corresponds to the splitting of the
leptons and quarks. The fermions follow the product representation of the two
algebras.

\medskip

\section{Observables in gravity and the spectral action}

\label{spectralaction}

The missing ingredient, in the above description of the Standard Model coupled
to gravity, is provided by a simple action principle---the spectral action
principle \cite{cc1}, \cite{cc2}, \cite{cc3}, \cite{cc4}---- that has the
geometric meaning of ``pure gravity" and delivers the action functional of the
Standard Model coupled to gravity when evaluated on $M\times F$. The spectral
action principle is the simple statement that the physical action is
determined by the spectrum of the Dirac operator $D$. The additivity of the
action forces it to be of the form $\mathrm{Trace}\,f\left(  D/\Lambda\right)
. $ This principle has now been tested in many interesting models including (\cf  \cite{chams}, \cite{Ioch},   \cite{Gayral},
\cite{Wulk1},   \cite{cc7},   \cite{cc3},   \cite{Broek},
  \cite{Stef})

\begin{itemize}

\item Superstring theory

\item noncommutative tori

\item Moyal planes

\item 4D-Moyal space

\item manifolds with boundary

\item in the presence of dilatons

\item for supersymmetric models

\item in the presence of torsion
\end{itemize}

To this action principle we want to apply the criterion of \emph{simplicity}
rather than that of \emph{beauty} given the relative nature of the latter.
Thus we imagine trying to explain this action principle to a Neanderthal man.
The spectral action principle, described below, passes the ``Neanderthal
test", since it amounts to counting spectral lines.

\smallskip The starting point at the conceptual level is the discussion of
observables in gravity. By the principle of gauge invariance the only
quantities which have a chance to be observable in gravity are those which are
invariant under the group of diffeomorphisms of the space-time $M$. Assuming
first that we deal with a classical manifold (and Wick rotate to Euclidean
signature for simplicity), one can form a number of such invariants (under
suitable convergence conditions) as the integrals of the form
\begin{equation}
\label{gravobs}\int_{M}\,F(K)\,\sqrt{g}\,d^{4}x
\end{equation}
where $F(K)$ is a scalar invariant function\footnote{the scalar curvature is
one example of such a function but there are many others} of the Riemann
curvature $K$. There are\footnote{See S.~Giddings, D.~Marolf, J.~Hartle,
\emph{Observables in effective gravity}, Phys.\ Rev.\  D {\bf 74}, 064018 (2006).} other more
complicated examples of such invariants, where those of the form
\eqref{gravobs} appear as the \emph{single integral} observables i.e. those
which add up when evaluated on the direct sum of geometric spaces. Now while
in theory a quantity like \eqref{gravobs} is observable it is almost
impossible to evaluate since it involves the knowledge of the entire
space-time and is in that way highly non localized. On the other hand,
spectral data  are available in localized form anywhere, and are
(asymptotically) of the form \eqref{gravobs} when they are of the additive
form
\begin{equation}
\mathrm{Trace}\,(f(D/\Lambda)),\label{specact}%
\end{equation}
where $D$ is the Dirac operator and $f$ is a positive even function of the
real variable while the parameter $\Lambda$ fixes the mass scale.  The
spectral action principle asserts that the fundamental action functional $S$
that allows to compare different geometric spaces at the classical level and
is used in the functional integration to go to the quantum level, is itself of
the form \eqref{specact}. The detailed form of the function $f$ is largely
irrelevant since the spectral action \eqref{specact} can be expanded in
decreasing powers of the scale $\Lambda$ and the function $f$ only appears
through the scalars
\begin{equation}
f_{k}=\,\,\int_{0}^{\infty}f(v)\,v^{k-1}\,dv.\label{coeff}%
\end{equation}

\medskip

\begin{table}[ptb]
\begin{center}
\medskip
\par%
\begin{tabular}
[c]{|c||c|}\hline
Standard Model & Spectral Action\\\hline
& \\
Higgs Boson & Inner metric$^{(0,1)}$\\
& \\\hline
& \\
Gauge bosons
\index{gauge!bosons}
& Inner metric$^{(1,0)}$\\
& \\\hline
& \\
Fermion masses & Dirac$^{(0,1)}$ in $\uparrow$\\
$u,\nu$ & \\\hline
& \\
CKM matrix & Dirac$^{(0,1)}$ in $\downarrow3)$\\
Masses down & \\\hline
& \\
Lepton mixing & Dirac$^{(0,1)}$ in $\downarrow1)$\\
Masses leptons $e$ & \\\hline
& \\
Majorana & Dirac$^{(0,1)}$ on\\
mass matrix & $E_{R}\oplus J_{F} E_{R}$\\\hline
& \\
Gauge couplings & Fixed at\\
& unification\\\hline
& \\
Higgs scattering & Fixed at\\
parameter & unification\\\hline
& \\
Tadpole constant & $- \mu_{0}^{2}\, |\mathbf{H}|^{2}$\\
& \\\hline
& \\
Graviton & Dirac$^{(1,0)}$\\
& \\\hline
\end{tabular}
\bigskip\bigskip
\end{center}
\caption{Conversion from Spectral Action to Standard Model}%
\label{smtospec}%
\end{table}

\smallskip As explained above the gauge potentials make good sense in the
framework of noncommutative geometry and come from the inner fluctuations of
the metric.

Let $M$ be a Riemannian spin $4$-manifold and $F$ \ the finite noncommutative
geometry of $KO$-dimension $6$ described above. Let $M\times F$ \ be endowed
with the product metric. Then by \cite{mc2}

\begin{enumerate}
\item The unimodular subgroup of the unitary group acting by the adjoint
representation $\mathrm{Ad}(u)$ in $\mathcal{H}$ is the group of gauge
transformations of SM.

\item The unimodular inner fluctuations of the metric give the gauge bosons of SM.

\item The full standard model (with neutrino mixing and seesaw mechanism)
minimally coupled to Einstein gravity is given in Euclidean form by the action
functional
\[
S=\,\mathrm{Tr}(f(D_{A}/\Lambda))+\frac12\,\langle\,J\,\tilde\xi,D_{A}%
\,\tilde\xi\rangle\,,\quad\tilde\xi\in\mathcal{H}^{+}_{cl} ,
\]
where $D_{A}$ is the Dirac operator with the unimodular inner fluctuations.
\end{enumerate}

The change of variables from the standard model to the spectral model is
summarized in Table \ref{smtospec}. We refer to \cite{mc2} for the notations.
To explain the table we note that the Higgs doublet corresponds to the inner
fluctuations of the Dirac operator along the discrete directions connecting
the right-handed and left-handed sectors of the quaternionic algebra. The
$SU(3)\times SU(2)\times U(1)$ gauge bosons are the inner fluctuations of the
Dirac operator along the continuous directions. The fermion masses, CKM mass
matrix and the Majorana mass matrices are all components of the Dirac operator
in discrete space.

\medskip

\section{Predictions}

\label{predsect}

The above spectral model can be used to make predictions assuming the ``big
desert" (absence of new physics up to unification scale) together with the
validity of the spectral action as an effective action at unification scale.
While the big-desert hypothesis is totally improbable, a rough agreement with
experiment would be a good indication for the spectral model. To be cautious,
since the change of scales from the $100$-GeV scale to the unification scale
is of the order of $10^{14}$, making predictions is a bit like trying to guess
if there is a fly in a cup of tea by looking at the earth from another
planetary system. Moreover since we do not have a quantum theory it might seem
that the presence of the renormalization ambiguity precludes the possibility
to predict the values of physical constants, which are in fact not constant
but depend upon the energy scale $\Lambda$. In fact the renormalization group
gives differential equations which govern their dependence upon $\Lambda$. The
intuitive idea behind this equation is that one can move down i.e. lower the
value of $\Lambda$ to $\Lambda-d\Lambda$ by integrating over the modes of
vibrations which have their frequency in the given interval. For the three
coupling constants $g_{i}$ (or rather their square $\alpha_{i}$) which govern
the three forces (excluding gravity) of the Standard Model, their dependence
upon the scale shows that they while they are quite different at low scale, they become
comparable at scales of the order of $10^{15}$ GeV. This suggested long ago
the idea that physics might become simpler and ``unified" at scales (called
unification scales) of that order. In our case we make the hypothesis that the
full spectral action is actually valid at the unification scale and we use the
numerical values of the various couplings as boundary values for the
renormalization group flow.

We can now describe the predictions obtained by comparing the spectral model
with the standard model coupled to gravity. The status of ``predictions" in
the above spectral model is based on two hypothesis:

\begin{enumerate}
\item The model holds at unification scale

\item One neglects the new physics up to unification scale.
\end{enumerate}

\bigskip The spectrum of the fermionic particles, which is the number of
states in the Hilbert space per family is predicted to be $4^{2}=16$ which is
a consequence of the algebra of the discrete space being $M_{2}\left(
\mathbb{H}\right)  \oplus M_{4}\left(  \mathbb{C}\right)  .$ In addition the
surviving algebra consistent with the axioms of noncommutative geometry, in
particular the order one condition, is given by $\mathbb{C}\oplus
\mathbb{H}\oplus M_{3}\left(  \mathbb{C}\right)  $ which gives rise to the
gauge group of the standard model. A consequence of this is that the $16$
spinors get the correct quantum number with respect to the standard model
gauge group which follows the decomposition:%
\begin{align*}
\left(  4,4\right)   &  \rightarrow\left(  1_{R}+1_{R}^{\prime}+2_{L}%
,1+3\right) \\
&  =\left(  1_{R},1\right)  +\left(  1_{R}^{\prime},1\right)  +\left(
2_{L},1\right) \\
&  +\left(  1_{R},3\right)  +\left(  1_{R}^{\prime},3\right)  +\left(
2_{L},3\right)
\end{align*}
These spinors correspond to $\nu_{R},$ $e_{R},$ $l_{L},$ $u_{R},$ $d_{R},$
$q_{L}$ respectively, where $l_{L}$ is the left-handed neutrino-electron
doublet and $q_{L}$ is the left-handed up-down quark doublet. In addition to
the gauge bosons of $SU(3)\times SU(2)\times U(1)$ which are the inner
fluctuations of the metric along continuous directions, we also have a Higgs
doublet which correspond to the inner fluctuations of the metric along the
discrete directions. What is peculiar about this Higgs doublet, is that its
mass term as determined from the spectral action comes with a negative sign
and a quartic term with a plus sign, thus predicting the phenomena of
spontaneous breakdown of the electroweak symmetry.

The value of the scale where the spectral action holds is the unification
scale since the spectral action delivers the same equality $g_{3}^{2}%
=g_{2}^{2}=\frac{5}{3}\,g_{1}^{2}$ which is common to all \textquotedblleft
Grand-Unified" theories. It gives more precisely the following unification of
the three gauge couplings:
\[
\frac{g_{3}^{2}\,f_{0}}{2\pi^{2}}=\frac{1}{4},\ \ \ \ \ \ g_{3}^{2}=g_{2}%
^{2}=\frac{5}{3}\,g_{1}^{2}\,.
\]
Here $f_{0}=f(0)$ is the value of the test function $f$ at $0$.

\smallskip The second feature which is predicted by the spectral model is that
one has a see-saw mechanism for neutrino masses with large $M_{R}\sim\Lambda$.

\smallskip

The third prediction that one gets by making the conversion from the spectral
model to the standard model is that the mass matrices satisfy the following
constraint at unification scale:
\[
\sum_{\sigma}\,(m_{\nu}^{\sigma})^{2}+(m_{e}^{\sigma})^{2}+3\,(m_{u}^{\sigma
})^{2}+3\,(m_{d}^{\sigma})^{2}=\,8\,M_{W}^{2}%
\]
In fact it is better to formulate this relation using the following quadratic
form in the Yukawa couplings:
\[
Y_{2}=\sum_{\sigma}\,(y_{\nu}^{\sigma})^{2}+(y_{e}^{\sigma})^{2}%
+3\,(y_{u}^{\sigma})^{2}+3\,(y_{d}^{\sigma})^{2}%
\]
so that the above prediction means that
\[
Y_{2}(S)=4\,g^{2}.
\]
This, using the renormalization group to compute the effective value at our
scale, yields a value of the top mass which is $1.04$ times the observed value
when neglecting the Yukawa couplings of the bottom quarks etc...and is hence
compatible with experiment.
We stress that the relations between the gauge coupling constants, and the
RG equations are carried for the interactions obtained by  assuming that the
spectral function is a cut-off function, and thus suppressing all higher
order terms. In a future work we shall show that if the spectral function $F\left(  D^{2}/\Lambda^2\right)  $
deviates by small perturbations from the cut-off function, higher order interactions can lead to small corrections which
alter the running of each of the gauge coupling constants and that their merging at a unification
scale is possible.  Thus it is worthwhile to study the case where the spectral function  deviates from the cut-off function. In addition, a distinctive feature of the spectral action is that the Higgs coupling is
proportional to the gauge couplings which yields a restriction on its mass.
If one naively solves the equation numerically, using the cut-off function, one gets a Higgs mass of the order
 of $170$ Gev. However, this answer is very sensitive to the value of the unification scale and to deviations of the spectral function from the cut-off function which will have substantial consequences. Therefore the mass of the Higgs can be different from the naive value of  $170$ Gev
 which is experimentally ruled out. The actual value can only be determined after working out the higher order corrections and including them in the RG equations.

On the other hand, the mass of the top quark is governed by the top quark
Yukawa coupling $k^{t}$ through the equation
\begin{equation}
m_{\mathrm{top}}(t)=\frac{1}{\sqrt{2}}\frac{2M}{g}\,k^{t}=\frac{1}{\sqrt{2}%
}\,v\,k^{t},
\end{equation}
 where {$v=\frac{2M}{g}$ }is the vacuum expectation value of the Higgs field.
\ All fermions get their masses by coupling to the Higgs through interactions
of the form {
\begin{equation}
kH\overline{\psi}\psi
\end{equation}
} After normalizing the kinetic energy of the Higgs field through the
redefinition $H\rightarrow${$\frac{\pi}{\sqrt{aF_{0}}}H,$} the mass term
becomes
\begin{equation}
\frac{\pi}{\sqrt{F_{0}}}\frac{k}{\sqrt{a}}H\overline{\psi}\psi
\end{equation}
and we notice that $%
{\displaystyle\sum\limits_{i}}
\left(  \frac{k_{i}}{\sqrt{a}}\right)  ^{2}=1.$ This gives a relation among
the fermions masses and the W- mass
\begin{equation}
{\sum_{\mathrm{generations}}m_{e}^{2}+m_{\nu}^{2}+3m_{d}^{2}+3m_{u}^{2}%
=8M_{W}^{2}.}%
\end{equation}
If the value of $g$ at a unification scale of {$10^{17}$ Gev} is taken to be
{$\sim0.517$ and }neglecting the $\tau$ neutrino Yukawa coupling, we get {%
\begin{equation}
k^{t}=\,\frac{2}{\sqrt{3}}\,g\sim0.597\,.
\end{equation}
} The numerical integration of the differential equation gives a top quark
mass of the order of $\,179$ Gev, and the agreement with experiment becomes
quite good if one takes into account the Yukawa coupling for neutrinos as
explained in details in \cite{mc2}. This indicates that the top quark mass is
less sensitive than the Higgs mass to the unification scale ambiguities. This
could be related to the fact that the fermionic action is much simpler than
the bosonic one which is only determined by an infinite expansion whose
reliability depends on the convergence of the higher order terms.

We also pass a severe test which gives confidence in the use of the spectral
action in determining the dynamics of all the fields appearing in the theory.
It is known that for the Einstein-Hilbert action to be consistent on manifolds
with boundary, a surface term must be added. This term is the integral of the
extrinsic curvature on the boundary of the manifold, and comes with a fixed
coefficient and sign. The spectral action when formulated on a space with
boundary gives exactly the Einstein-Hilbert action with the boundary term
correct in both coefficient and sign. This is a strong test because if the
Laplacian operator is used instead of the square of the Dirac operator, then
an inconsistent answer is obtained \cite{cc7}.

Finally, we note that the scale $\Lambda$ appears as a free parameter in the
spectral action. It is more natural if it can arise as the vev of a dynamical
field. We thus introduce the dilaton field $\phi$ and replace the operator
$D^{2}$ in the spectral action by
\[
P=e^{-\phi}D^{2}e^{-\phi}%
\]
A shift in the dilaton field $\phi\rightarrow\phi+\ln\Lambda$ transforms
$P\rightarrow\frac{1}{\Lambda^{2}}P.$ This implies that the physical metric
gets rescaled, and so does the Higgs field according to $H^{\prime}=He^{-\phi
}.$ Although the Higgs fields $H$ gets a vev of the order of the Planck scale,
however, the physical field $H^{\prime}$ has its vev suppressed through the
dilaton coupling $e^{-\phi}$. Thus if $\left\langle \phi\right\rangle \sim40$
in Planck units, then $e^{-\phi}\sim10^{-19}$. Thus the problem of explaining
the very low mass scale of fermion masses reduces to explaining the origin of
a dilaton vev of the order of $10^{2}$ (\cf \cite{cc3}).

\begin{figure}
\begin{center}
\includegraphics[scale=1.2]{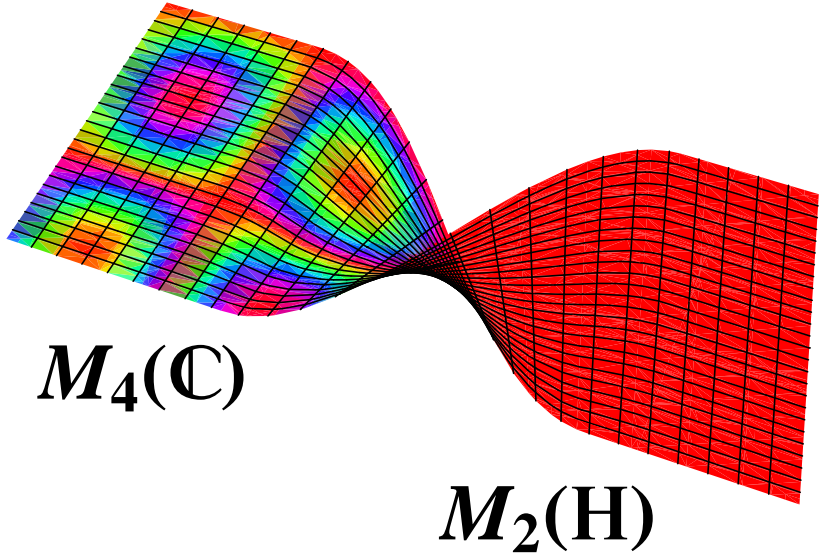}
\end{center}
\caption{The ``color" side and the ``quaternionic" (chiral) side of the spectral model.\label{double} }
\end{figure}

It may be helpful to draw a geometrical picture (\cf Figure \ref{double})
of the emerging space-time as
dictated by noncommutative geometry. We can imagine that space-time is a
parallel universe where each copy is a four-dimensional manifold. On one
universe the algebra is the algebra of $2\times2$ quaternionic matrices,
broken by the chirality operator to $\left(  \mathbb{C}\oplus\mathbb{C}%
^{\prime}\right)  _{R}\mathbb{\ }\oplus\mathbb{H}_{L}$. This is the
quaternionic universe. On the other ``color" universe we have the algebra of $4\times4$ complex
matrices which are decomposed into $1\times1$ and $3\times3$ matrices
providing the split between leptons and quarks. This is similar in spirit to
the Pati-Salam model who considered the lepton number to be the fourth color.
We can refer to this as the color universe. The fermions live on both
universes and thus have the tensor product representation. The Higgs doublet
is the field that connects the right to the left sectors in the quaternionic
universe, and this joining will provide masses to the quarks and leptons. To
take care of the reality condition, both a spinor and its conjugate are
present. The KO dimension of $6$ for the discrete space implies that the
conjugate spinor is not independent of the spinor, but is related to it. The
mixing between the neutral spin state and its conjugate is done through a
neutral scalar field, responsible for the see-saw mechanism that gives the
left-handed neutrino a tiny mass.

\section{Cutoff scale and the spectral approach}

There is one important advantage of the spectral point of view when compared to the
old idea of a discrete space-time, which is that continuous symmetry groups survive the operation of truncating the Hilbert space $\cH$ to the finite dimensional subspace $\cH(\Lambda)$ corresponding to eigenvectors of the Dirac operator for eigenvalues $\leq \Lambda$
where $\Lambda$ is a cutoff scale. Indeed, any unitary operator $U$ commuting  with $D$
will automatically restrict to $\cH(\Lambda)$. It could well be that the coherence of the spectral action principle indicates that our continuum picture of space-time is only an approximation to a completely {\em finite} spectral geometry whose underlying Hilbert space is {\em finite dimensional}. Of course the basic ingredients such as $J$ and $\gamma$ will still be present, but the algebra $\cA$ itself will have no reason to remain commutative. In this scenario, once we go up in energy towards the unification scale, the small amount of noncommutativity encoded in the finite geometry $F$ to model the present scale, will gradually creep  in and invade the whole algebra of coordinates which will become a huge matrix algebra at Planck scale.  The noncommutativity of the algebra of coordinates means that the ``internal" degrees of freedom have gradually replaced the external ones and that the notion of ``point" has disappeared since a matrix algebra admits only one irreducible representation.

\smallskip

Going backwards from this unified picture down to our scale, this raises in particular the possibility that  geometry   only
emerges after a suitable symmetry breaking mechanism which extends
to the full gravitational sector the electroweak symmetry breaking.
The invariance of the spectral action under
the symplectic unitary group in Hilbert space is broken during this
process  to the compact group of isometries of a given geometry.

\smallskip

One basic issue in trying to give substance and test the above scenario is to go back from the Euclidean signature that we have been using throughout for simplicity to the usual Minkowski signature. One possible approach towards this is contained in our paper \cite{cc8} in which the spectral action is used to analyze the static situation obtained as the product of a three geometry by a small circle of length $\beta=\frac{1}{kT}$.

\section*{Acknowledgement}
The work of AHC is supported in part by the National Science Foundation grant Phys-0854779.

\end{document}